\documentclass[seceq,preprint]{ptptex}

\usepackage{graphicx}

\def\ds{\displaystyle}
\def\cP{{\cal P}}
\def\Re{{\rm Re}}

\newcommand{\bn}{\mbox{\boldmath{$n$}}}

\newcommand{\bbs}{\mbox{\boldmath{$s$}}}

\newcommand{\bzero}{\mbox{\boldmath{$0$}}}

\newcommand{\bbr}{\mbox{\boldmath{$r$}}}
\newcommand{\bt}{\mbox{\boldmath{$t$}}}
\newcommand{\bbt}{\mbox{\boldmath{$t$}}}
\title{%        %You can use \\ for explicit line-break
Exact Analytic Continuation  with Respect to  the Replica Number
in the Discrete Random Energy Model
of Finite System Size
}

\author{%       %Use \scshape  for the family name
Kenzo \textsc{Ogure}$^{1,}$\footnote{E-mail: ogure@icrr.u-tokyo.ac.jp} 
and Yoshiyuki \textsc{Kabashima}$^{2,}$\footnote{
E-mail: kaba@dis.titech.ac.jp}
}

\inst{%         %Affiliation, neglected when [addenda] or [errata]
$^1$Theory group, Institute for Cosmic Ray Research, Kashiwano-ha 5-1-5, Kashiwa 2778582, Japan\\
$^2$Department of Computational Intelligence and Systems Science, 
Tokyo Institute of Technology, Yokohama 2268502, Japan

}
\abst{%  %this abstract is neglected when [addenda] or [errata]
An expression for the moment of partition function 
valid for any finite system size $N$ and complex 
power $n\ (\Re(n)>0)$ is obtained for a simple spin glass model 
termed the {\em discrete random energy model} (DREM).
We investigate the behavior of the moment in the thermodynamic 
limit $N \to \infty$ using this expression, and 
find that a phase transition occurs at a certain real replica number
when the temperature is sufficiently low, directly clarifying
the scenario of replica symmetry breaking of DREM in 
the replica number space {\em without using the replica trick}. 
The validity of the expression is numerically confirmed. 
}

\begin{document}

\maketitle

\section{Introduction}
The replica method (RM) is one of the few analytic schemes available 
for the research of disordered systems \cite{Beyond}. 
In physics, this method has been well known since the 1970s 
and has been successfully applied to the analyses of spin-glass 
models \cite{EA,SK,Parisi}, although
the essential idea behind the method can be dated back to 
the end of 1920s where it appeared as a theorem 
for computing the average of logarithms \cite{Hardy,Riesz,Hardy_book}.  
More recently, considerable attention has been paid to
the similarity between statistical mechanics of 
disordered systems and the Bayesian method 
in problems related to information processing (IP) \cite{Nishimori}.  
The number and variety of applications of RM to problems 
the IP research are increasing rapidly, 
including error-correcting codes \cite{Sourlas,KS}, image 
restoration \cite{NishimoriWong,TanakaKazu}, 
neural networks \cite{learning}, 
combinatorial problems \cite{Ksat,Napsak}, and so on. 

Although only the limit value 
$(1/N) \left \langle \ln Z \right \rangle=\lim_{n \to 0} 
\left (\left \langle Z^n \right \rangle^{1/N} -1 \right )/n$
is usually emphasized, RM can be considered to be a systematic 
procedure for calculating generalized moments 
$\left \langle Z^n \right \rangle$ of the partition function $Z$ 
in the case of $N \to \infty$ when $Z$ depends on a certain 
external randomness. Here, $N$ characterizes the system size, 
$n \in {\bf R} \mbox{ (or {\bf C})}$ is a real (or complex\cite{Saakian})
number and $\left \langle  \cdots \right \rangle$ means 
the average over the external randomness. 
For most problems, a direct assessment of 
$\left \langle Z^n \right \rangle$
is difficult for a general $n \in {\bf R} \mbox{ (or {\bf C})}$, 
whereas such an assessment for natural numbers 
$n=1,2,\ldots $ is possible in the thermodynamic 
limit $N \to \infty$. Therefore, $\left \langle Z^n \right \rangle$ is 
first computed for natural numbers and their analytic
continuation is used to extend $\left \langle Z^n \right \rangle$
to $n \in {\bf R} \mbox{ (or {\bf C})}$. 
This is usually termed the {\em replica trick}.

However, the validity of the replica trick is doubtful. 
The most obvious analytic continuation, 
obtained under the replica symmetric (RS) ansatz, 
sometimes leads to the wrong results.
The causes of these errors were actively debated 
in the 1970s until Parisi discovered 
the replica-symmetry-breaking (RSB) scheme 
for constructing reasonable solutions within the framework of RM 
\cite{Parisi}. 
Since this discovery, there have been no known examples 
for which physically wrong results have been derived by RM, 
in conjunction with the Parisi scheme if necessary.  
Therefore, RM is now empirically 
recognized as a reliable procedure in physics, although the mathematical 
justification of the replica trick still remains open. 
However, this problem is now generating interest
again, in particular, in the application of RM to IP problems.
Ths is because most theories in IP research have 
conventionally been developed with mathematical 
soundness highly valued\cite{Cover,Learning}. 

The purpose of this paper is to provide a method
to approach the problems of RM. 
Specifically, we give a useful formula to
compute $\left \langle Z^n \right \rangle$ {\em directly}
for $n \in {\bf C}$ at {\em finite} $N$
for a simple spin glass model, termed the discrete 
random energy model (DREM) \cite{REM,Mou1,Mou2}. 
This formula is numerically tractable, so 
one can directly observe how the system approaches 
the thermodynamic limit with the aid of numerical
calculation. Furthermore, this {\em analytically} clarifies 
the correct behavior for $N \to \infty$, making 
a direct examination of the validity of RM. 

We have two main reasons for picking DREM from 
among the various disordered systems. 
First, this model is simple enough to handle analytically. 
It is already known that RM, in conjunction with 
the Parisi scheme, can evaluate the correct free energy for 
a family of random energy models (REM) including DREM 
at the limit of $n \to 0$ \cite{REM}. 
However, the existing procedure seems at odds with a theorem 
concerning analytic continuation provided by 
Carlson \cite{Carlson,van_Hemmen}, which holds for DREM 
of finite $N$ claiming the uniqueness of analytic 
continuation from natural numbers $n \in {\bf N}$ to complex 
numbers $n \in {\bf C}$, when the temperature is sufficiently low.
For this, our approach shows that a phase transition 
occurs at a certain critical replica number $n_c \in [0,1]$ 
in such cases, which clarifies that RM can be consistent 
with Carlson's theorem. 
This may offer a useful discipline 
to perform analytic continuation 
from $n \in {\bf N}$ to $n \in {\bf R} \mbox{ (or \bf C)}$ in RM. 
The second reason is a relationship between REM and certain problems of IP. 
Recent research on error-correcting codes has revealed that 
REM is closely related to a randomly constructed code \cite{Sourlas,KS}. 
These codes are known to provide the best error correction performance 
in information theory\cite{Shannon}, and 
the performance evaluation of such codes is similar to
the computation of $\left \langle Z^n \right 
\rangle$ for $n \in {\bf R}$ 
\cite{Reliability} (see appendix \ref{ECC}). 
Therefore, the current investigation will indirectly
justify the RM-based analysis of error-correcting codes 
performed previously\cite{KS,KMS,JPAspecial_issue}.

This paper is organized as follows. 
In section \ref{replica} we introduce DREM and briefly 
review how RM has been employed in conventional analysis of this system. 
Referring to Carlson's theorem, we address how 
the conventional scenario for taking a limit $n \to 0$ 
seems controversial. In order to resolve this difficulty, 
we propose in section \ref{exact} a new scheme to directly evaluate 
$\left \langle Z^n \right \rangle$ for REM 
of finite $N$ and complex 
$n$ without using the replica trick. 
Taking the limit $N \to \infty$, we analytically clarify 
how $\left \langle Z^n \right \rangle$ behaves in the thermodynamic limit
and numerically verify this behavior. 
In section \ref{thermo} we show how RM can be consistent with 
the results obtained by the proposed scheme. 
Section \ref{summary} is a summary.

\section{Replica method in the discrete 
random energy model (DREM)}\label{replica}
In order to clearly state the problem addressed in this paper, 
we first review how RM has been conventionally employed 
in analyzing REM \cite{REM,Mezard_Extreme}. 
For convenience in the later analysis, we mainly concentrate 
on DREM, but the addressed problem is widely shared with other 
versions of REM as well. 

A DREM is composed of $2^N$ states, the energies of which, $\epsilon_A$
$(A =1,2,\ldots, 2^N)$, are independently drawn from 
an identical distribution 
\begin{eqnarray}
P(E_i)=2^{-M}
\left(
\begin{array}{c}
M\\
\frac{1}{2}M+E_i
\end{array}
\right),\ \ 
\left (E_i=i-\frac{M}{2} \right ), 
\label{prob}
\end{eqnarray}
over $M+1$ energy levels 
%%%%%% $E_i=0,1,\ldots,M$. revised by YK 04/03/01. 
$E_i=-M/2,-M/2+1, \ldots, M/2-1, M/2$. 
For each realization $\{ \epsilon_A\}$, the partition function 
\begin{eqnarray}
Z=\sum_{A=1}^{2^N}\exp(-\beta \epsilon _A), 
\label{part}
\end{eqnarray}
and the free energy (density)
\begin{eqnarray}
F=-\frac{kT}{N}\log{Z}
\label{free_energy}
\end{eqnarray}
can be used for computing various thermal averages. 
However, when the {\em configurational average} 
is required, one has to compute the averaged free energy
$\left\langle F \right\rangle=
-\frac{kT}{N}\left\langle\log{Z}\right\rangle$,
the direct evaluation of which is generally difficult. 
Here, $\left \langle \cdots \right \rangle$ denotes 
the configurational average with respect to $\{ \epsilon_A\}$. 
On the other hand, the moment of the partition function 
$\left\langle Z^n \right\rangle$ can be easily calculated 
in various models for natural numbers $n=1,2,\ldots$. 
Therefore, the replica method evaluates the averaged free energy 
using the {\em replica trick}
\begin{eqnarray}
\frac{1}{N}\left \langle \log{Z}\right \rangle =\lim_{n\rightarrow 0}
\frac{\left\langle Z^n \right\rangle^{\frac{1}{N}}-1}{n}, 
\label{repid}
\end{eqnarray}
analytically continuting the expression of 
$\left\langle Z^n \right\rangle$
for $n=1,2,\ldots$ to that for real (or complex) numbers $n$.

For a given natural number $n$, the moment of DREM is calculated as
\begin{eqnarray}
\left \langle Z^n \right \rangle
&=&
\sum_{A_1=1}^{2^N}
\sum_{A_2=1}^{2^N}
\dots
\sum_{A_n=1}^{2^N}
\sum_{i_1=0}^{M}P \left (E_{i_1}^{(1)} \right )
\sum_{i_2=0}^{M}P \left (E_{i_2}^{(2)} \right )
\dots
\sum_{i_{2^N}=0}^{M}P \left (E_{i_{2^N}}^{(2^N)} \right )\nonumber\\
&&\exp \left ( 
-\beta \sum_{B=1}^{2^N}E _{i_B}^{(B)}\sum_{\mu =1}^n\delta _{BA_{\mu}}
\right )\cr
&=& 
e^{\frac{n M\beta }{2}} \sum_{A_1=1}^{2^N}
\sum_{A_2=1}^{2^N}
\dots
\sum_{A_n=1}^{2^N}
\prod_{B=1}^{2^N} 
\left ( \frac{1+e^{-\beta \sum_{\mu =1}^n \delta_{BA_\mu}}}{2}
\right )^M \cr
&=&
\sum_{A_1=1}^{2^N}
\sum_{A_2=1}^{2^N}
\dots
\sum_{A_n=1}^{2^N}
\prod_{B=1}^{2^N} 
\exp \left (N \alpha I \left (\beta \sum_{\mu=1}^n  \delta_{BA_\mu}
\right )
\right )
\label{mom}
\end{eqnarray}
where 
$\alpha=M/N$, the function $I(x)$ is defined as
\begin{eqnarray}
I(x)= \log \left (\cosh \frac{x}{2} \right ). 
\label{I}
\end{eqnarray}
An identity $\prod_{B=1}^{2^N} \exp \left (-\frac{\beta}{2} 
\sum_{\mu=1}^n \delta_{BA_\mu} \right ) =\exp \left 
(-\frac{n\beta}{2} \right )$ was employed
for obtaining the final expression of Eq. (\ref{mom}). 
From now on, we focus on the case $\alpha >1$, 
for which the replica symmetry can be broken when the temperature is 
sufficiently low.

Unfortunately, performing the summation 
in Eq.~(\ref{mom}) exactly is difficult. 
Instead, in conventional RM, Eq.~(\ref{mom}) is represented 
by the most dominant contribution in the summation.  
This can be justified for natural numbers $n$ in the limit $N \to \infty$. 
Notice that the summation is invariant with respect to the permutation of 
the replica indices $\mu=1,2,\ldots,n$. 
This {\em replica symmetry} restricts the candidates
of the most dominant contribution to three possibilities, 
which are here referred to as solutions of the replica symmetric 
1 (RS1), the replica symmetric 2 
(RS2), and the 1-step replica symmetry breaking (1RSB). 

\begin{itemize}
\item RS1\\
In RS1, all of the $n$ replicas are assumed to 
occupy $n$ different states $B(=1,2,\ldots, 2^N)$. 
Therefore, for a given $B$, 
\begin{eqnarray}
\sum_{\mu=1}^n \delta_{BA_{\mu}} =
\left\{
\begin{array}{ll}
1 & (\mbox{when $B$ is one of the $n$ occupied states}), \\
0 & (\mbox{otherwise}).
\end{array}
\right.
\end{eqnarray}
The number of ways to assign $n$ replicas to
$n$ out of $2^N$ different states 
is 
\begin{eqnarray}
2^N\times(2^N-1)\times\dots \times(2^N-n+1) \sim 2^{nN}. 
\end{eqnarray}
For each case, the configuration of this type contributes
\begin{eqnarray}
\exp{ \left (nN \alpha {I}(\beta) \right )}, 
\end{eqnarray}
in Eq.~(\ref{mom}). This means that the contribution to 
the moment from RS1 becomes 
\begin{eqnarray}
\left\langle Z^n \right\rangle
&=&
2^{nN}\exp{\left (nN \alpha {I}(\beta) \right )}\cr
&=&
\exp{nN \left (\log{2}+ 
\alpha \log{ \left (\cosh{\frac{\beta}{2}} \right )
} \right )}.\label{rs1}
\end{eqnarray}

\item RS2\\
In RS2, all of the $n$ replicas are assumed 
to occupy a particular state $B$. 
Therefore, for a given $B$, 
\begin{eqnarray}
\sum_{\mu=1}^n \delta_{BA_{\mu}}
=
\left\{
\begin{array}{ll}
n & (\mbox{when $B$ is the occupied state}),\\
0 & (\mbox{otherwise}).
\end{array}
\right.
\end{eqnarray}
The number of ways to choose one out of $2^N$ states is $2^N$. 
For each case, the configuration of this type contributes 
\begin{eqnarray}
\exp{ \left (N \alpha {I}(n\beta) \right )}, 
\end{eqnarray}
in Eq.~(\ref{mom}), which indicates that 
the contribution from RS2 is 
\begin{eqnarray}
\left\langle Z^n\right\rangle
&=&
2^{N}\exp{ \left (nN\alpha {I}(n\beta) \right )} \cr
&=&
\exp{N \left (\log{2}+ \alpha 
\log{ \left (\cosh{\frac{n\beta}{2}} \right )} \right )}.
\label{rs2}
\end{eqnarray}

\item 1RSB\\
In 1RSB, $n$ replicas are assumed to be equally assigned 
to $n/m $ states $B$, 
where $m$ is an aliquot of $n$. 
Therefore, for a given $B$, 
\begin{eqnarray}
\sum_{\mu=1}^n \delta _{BA_{\mu}}
=
\left\{
\begin{array}{ll}
m &  (\mbox{when $B$ is one of the $n/m$ occupied states}),\\
0 & (\mbox{otherwise}).
\end{array}
\right.
\end{eqnarray}
The number of ways to select $n/m$ out of $2^N$ states 
equally assigning $n$ replicas to the $n/m$ 
states is 
\begin{eqnarray}
\frac{(2^N)!}{(2^N-n/m)! }\times \frac{n!}{m^{n/m}}
\sim
2^{\frac{n}{m}N}. 
\end{eqnarray}
For each case, the configuration of this type contributes
\begin{eqnarray}
\exp{\left (\frac{n}{m} N\alpha {I}(m\beta) \right )}, 
\end{eqnarray}
in Eq.~(\ref{mom}). 
Taking all the possibile values of $m$ into account, 
the contribution from 1RSB can be summarized as 
\begin{eqnarray}
\left\langle Z^n\right\rangle
&=&
\sum_{m}2^{ \frac{n}{m} N}
\exp \left (\frac{n}{m} N \alpha {I} (m\beta) \right )\cr
&=&
\sum_{m}\exp \left (\frac{n}{m} N 
\left [\log{2}+ \alpha {I}\left (m \beta \right ) \right ] \right )\cr 
&\sim&
\exp \left ({\mathop{\rm extr}_{m} \left
\{\frac{n}{m} N \left [\log{2}+ \alpha { I}
\left (m\beta  \right ) \right ] \right \}} \right ).
\end{eqnarray}
In the last expression, we have replaced the summation over $m$ with 
the extremization with respect to 
$m$ ($\mathop{\rm extr}_m \{ \cdots \}$), which 
is hopefully valid for large $N$, 
analytically continuing the expression with respect to $m$ 
from natural numbers to real numbers. 
The extremization with respect to $m$ yields the condition 
\begin{eqnarray}
\log{2}+ \alpha {I}
\left (m \beta  \right )=
\alpha m \beta {I}^\prime
\left (m \beta  \right ),\label{con}
\end{eqnarray}
which implies that the moment is expressed as 
\begin{eqnarray}
\left\langle Z^n \right\rangle
&=&
\exp \left [n N \alpha \beta {I}^\prime\left (m_c \beta  \right ) 
\right ], 
\end{eqnarray}
where $m_c$ is the solution of Eq.~(\ref{con}). 
Eq.~(\ref{I}) indicates that $m_c$ can be represented as
$m_c=\beta_c/\beta$, where 
the critical inverse tempereture 
$\beta_c>0 $ is determined by 
\begin{eqnarray}
\log{2}+\alpha \left (\log \left (\cosh \frac{\beta_c}{2} 
\right )-\frac{\beta_c}{2}\tanh\frac{\beta_c}{2}\right )=0. 
\label{cdef}
\end{eqnarray}
This provides a simple expression of the 1RSB solution as 
\begin{eqnarray}
\left \langle Z^n \right \rangle 
&=&
\exp\left (nN \alpha \beta I^\prime(\beta_c) \right) \cr
&=& \exp\left ( \frac{nN \alpha \beta}{2} 
\tanh \frac{\beta_c}{2} \right ). 
\label{1rsb}
\end{eqnarray}
\end{itemize}

\begin{figure}[t]
\centerline{\includegraphics[width=12cm] {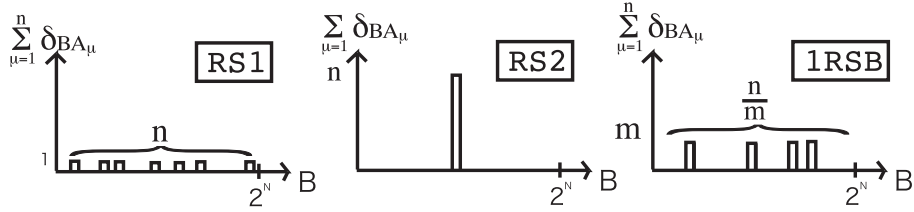}}
\caption{The configuration of replicas for each solution.  
In RS1, $n$ replicas are distributed in different states.  
In RS2, $n$ replicas are concentrated in one state.  In 1RSB, 
$n$ replicas are equally assigned to $n/m$ states.  
For 1RSB, however, this figure does not directly correspond 
to the solution because the critical value of $m$ is 
not neccesarily an integer and can be larger than 
$n$ for $\beta>\beta_c$.}
\label{RSFIG}
\end{figure}

The configurations of the replicas assumed for RS1, RS2, and 1RSB 
are pictorially
presented in Fig.~\ref{RSFIG}.
It should be emphasized here that the above three solutions
are derived for $n=1,2,\ldots$, assuming $N$ is sufficiently large. 
However, the obtained expressions are likely to hold for real 
$n$ as well. Therefore, in conventional analysis, the replica 
trick of Eq.~(\ref{repid}) is carried out, selecting one possibly relevant 
solution of the three, which is hopefully valid for large $N$.

\begin{figure}[t]
\centerline{\includegraphics[width=12cm]{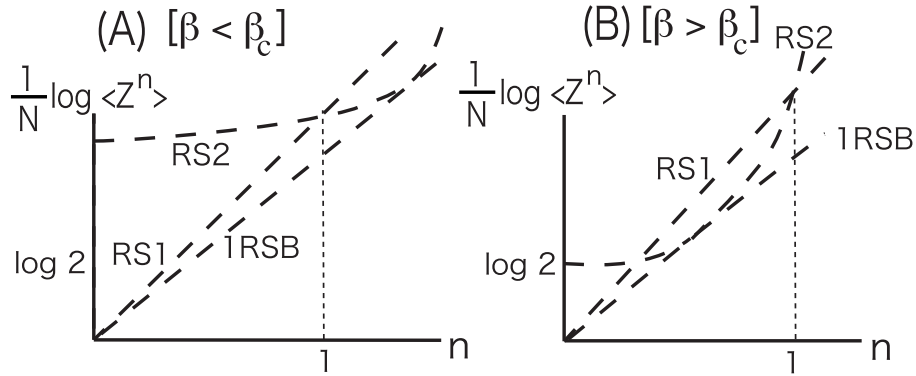}}
\caption{
Solutions obtained under the RS1, RS2 and 1RSB assumptions. 
RS1 and RS2 agree at $n=1$
while RS2 contacts 1RSB at a certain point of positive $n$. 
(a) For $\beta < \beta_c$, the contact 
point is placed in the region of $n > 1$.  
(b) For $\beta> \beta_c$, on the other hand, it is located in $0 < n < 1$.  
}
\label{phasecom}
\end{figure}

%%%%%%%%%
%%%%%%%%%
%%%%%%%%%
The existing prescription for selecting the relevant solution 
is as follows \cite{REM}: For small $n$, RS1, RS2, and 1RSB 
are ordered RS2 $>$ RS1 $>$ 1RSB from the viewpoint of 
their amplitudes (Fig. \ref{phasecom}). 
The contribution from RS2, however, 
converges to $2^N$ rather than unity for $n \to 0$, 
and therefore, the replica trick leads to divergence. 
Hence, this solution is discarded. After excluding this solution, 
the leading contribution always comes from RS1, 
which guarantees a finite limit in Eq.~(\ref{repid}).

%%%%%%%%%% revised by YK 04/03/01
Actually, the answer obtained from this solution is 
correct in the case of high temperatures $0 < \beta < \beta_c$. 
%where $\beta_c$ is an critical inverse temperature determined
%by the condition 
%\begin{eqnarray}
%\log{2}+\frac{1}{N}{\tilde I}(\beta_c)=
%\frac{1}{N}\beta_c{\tilde I}^\prime(\beta_c).
%\label{betacdef}
%\end{eqnarray}
However, this solution becomes invalid for low temperatures
$\beta>\beta_c$, for which the correct answer is 
provided by 1RSB. It may be worth noting that 
the value of $m_c$, in this low temperature case, is 
placed in the interval $n \le m_c \le 1$ when $n \to 0$, 
which is out of the ordinary range $1 \le m_c \le n$ for $n=1,2,\ldots$. 
%%%%%%%%%
%%%%%%%%%
%%%%%%%%%
%%The above 
This prescription for taking the $n \to 0$ limit
is empirically justified for a family of REMs because 
it reproduces the correct answers obtainable 
by other schemes at the limit of $n \to 0$\cite{REM}.
However, the following two issues still need further investigation. 

The first issue is the reason for expurgating RS2. 
Carlson's theorem, which guarantees the uniqueness 
of analytic continuation from natural $n \in {\bf N}$ to 
complex $n \in {\bf C}$, might be useful for resolving 
this problem\cite{Carlson,van_Hemmen}. 
Unlike other systems such as the 
Sherrington-Kirkpatrick model\cite{SK} and the original REM\cite{REM}, 
%%%%%%%
%%%%%%% revised by YK 04/03/01
%%%%%%%
%%%%%%% About Carlson's theorem
%%%%%%%
%%%%%%%
a modified moment of the current DREM 
$\left \langle (e^{-M\beta /2} Z)^n 
\right \rangle^{1/N}$, which is extended from $n \in {\bf N}$ 
to $n \in {\bf C}$, 
satisfies an inequality,  
\begin{eqnarray}
\left | 
\left \langle 
\left (e^{-M\beta/2} Z \right )^n 
\right \rangle^{1/N} 
\right |
&\le &
\left \langle \left (e^{-M\beta /2} Z \right )^{\Re(n)} 
\right \rangle^{1/N} \cr
&=& \left \langle \left (\sum_{A=1}^{2^N}
\exp\left [ -\beta (\epsilon_A +M/2) \right] \right )^{\Re(n)} 
\right \rangle^{1/N}  \cr
&\le&  \left \langle \left (\sum_{A=1}^{2^N} 1 \right )^{\Re(n)} \right \rangle^{1/N} = 2^{\Re (n)} < 
O\left (\exp \left [ \pi |n| \right ] \right ), 
\label{eq:necessary_carlson}
\end{eqnarray}
for $\Re(n) \ge 0$ and $\forall{N}=1,2,\ldots$, 
since $\epsilon_A +M/2$ is lower bounded by $0$. 
Suppose that we could construct another extention 
$\psi(n;N)$ $(n \in {\bf C})$, 
which satisfies the the growth condition, 
$\psi(n;N) < O\left (\exp \left [ \pi |n| \right ] \right )$
\footnote{This condition 
is necessary to exclude a trivial multiplicity caused by 
addition of certain analytic functions 
which vanish at all the natural numbers 
$n=1,2,\ldots$, such as $\sin(\pi n)$. },
and agrees with $\left \langle (e^{-M\beta /2} Z)^n 
\right \rangle^{1/N}$ at all the natural numbers $n=1,2,\ldots$. 
This indicates that a similar inequality 
$\left | \psi(n;N) - \left \langle (e^{-M\beta /2} Z)^n 
\right \rangle^{1/N} \right | \le 
\left | \psi(n;N) \right |+
\left | \left \langle (e^{-M\beta /2} Z)^n \right \rangle^{1/N} \right |
< O\left (\exp [\pi |n|] \right )$ 
holds for ${\rm Re}(n) \ge 0 $ 
and the difference 
$\left | \psi(n;N) -\left \langle (e^{-M\beta /2} Z)^n 
\right \rangle^{1/N} \right | $ vanishes
at all the natural numbers $n=1,2,\ldots$, 
as $\psi(n;N)$ and $\left \langle (e^{-M\beta /2} Z)^n 
\right \rangle^{1/N}$ completely coincide for $n \in {\bf N}$. 
Then, Carlson's theorem (Theorem 5.81 in page 186 of 
Ref. \citen{Carlson})
ensures that  
$\left | \psi(n;N)- 
\left \langle (e^{-M\beta /2} Z)^n \right \rangle^{1/N} \right | $
is identical to $0$, implying that $\psi(n;N)$ and 
$\left \langle (e^{-M\beta /2} Z)^n \right \rangle^{1/N}$ are 
identical and, therefore, analytic continuation of 
$\left \langle (e^{-M\beta /2} Z)^n \right \rangle^{1/N}$
from natural $n \in {\bf N}$ to complex $n \in {\bf C}$ 
can be uniquely determined, 
unless analyticity of $\left \langle (e^{-M\beta /2} Z)^n 
\right \rangle^{1/N}$ is lost in the limit $N \to \infty$. 
Since $e^{-M\beta/2}$ is a non-vanishing constant, 
this means that analytic continuation of the moment 
$\left \langle Z^n \right \rangle^{1/N}$ is also unique
as long as $\left \langle Z^n 
\right \rangle^{1/N}$ remains analytic with respect to 
$n$ in the limit $N \to \infty$.
%%%%%%% 
%%%%%%%
%%%%%%% About Carlson's theorem
%%%%%%%
%%%%%%%
Hence, the dominant solution for $n=1,2,\dots$, rather than for 
$0<n<1$, should be selected as the relevant solution for $n \to 0$. 
This recipe succesfully reproduces the correct answer for 
the high temperature region $0< \beta < \beta_c$. 
However, this is still not fully satisfactory because 
RS2 becomes dominant for $n=1,2,\ldots$ in the case 
of $ \beta > \beta_c$ and, therefore, should be selected as 
the relevant solution for $n \to 0$ if analyticity is preserved
for $N \to \infty$, which, unfortunately, leads to a wrong answer. 
Hence, a certain phase transition must occur at a critical 
replica number $0<n_c<1$ for $\beta > \beta_c$. 
However, to the knowledge of the authors, 
such a phase transition with respect to  $n$ 
has not yet been fully examined for most disordered 
systems\cite{van_Hemmen,Horiguchi}. This might be because so far 
the greatest attention 
has been paid only to the final results at the limit $n \to 0$. 
However, detailed analysis of phase transitions of this type 
may soon be needed as the replica calculation for 
non-vanishing $n$ has recently begun to be employed in problems 
related to IP\cite{Monasson,Reliability} 
%%% revised by YK 04/03/02. 
and analysis of certain dynamics 
involved with multiple time scales\cite{Coolen}. 

The second question is the origin of 1RSB. In conventional analysis 
at the limit $n \to 0$\cite{REM,Mou1,Mou2,Mezard_Extreme},
this solution is introduced by modifying RS1 to keep
the entropy of the correct solution 
non-negative for $\beta > \beta_c$. 
It would appear that 1RSB originates from RS1. 
However, at least for positive $n$, this association 
seems unlikely because the two solutions cross each other 
only at $n=0$ (see Fig. \ref{phasecom}). 
Therefore, it is impossible to relate
the origin of 1RSB to RS1 at large $n$, from which 
the solutions of smaller $n$ are extrapolated. 
On the other hand, 1RSB contacts RS2 
at a certain point of positive $n$, 
implying that 1RSB bifurcates from RS2. 
As the contact point is located in $0<n<1$
for $\beta > \beta_c$, 
this seems consistent with the aforementioned possible 
phase trainsition at $n=n_c$ (see Fig. \ref{phasecom} (b)).
Nevertheless, such a scenario cannot be so easily 
accepted since RS2 still dominates 1RSB even below the contact point; 
a mere contact does not change the dominance
between the two solutions.

It may help in resolving these problems to 
analyze DREM employing a completely different methodology. 
In the next section, we provide a scheme to calculate 
the moment of DREM as a step towards 
clarifying the mysteries of RM. The proposed 
method is powerful enough to 
evaluate the moment in the right half of the complex 
%plain revised by YK 04/03/01 
plane 
$\Re(n)>0$ 
for any arbitrary finite $N$ without using 
the replica trick, which supplies a model answer for RM. 

%%%%%%%
%%%%%%%
%%%%%%%
%%%%%%% 2003/10/20 kokomade
%%%%%%%
%%%%%%%
%%%%%%%

\section{Direct calculation of the moment for grand canonical DREM 
(GCDREM)}\label{exact}
\subsection{General formula}
In order to introduce a novel scheme for caluculating moments of 
the partition function of DREM, we first rewrite the partition 
function using the occupation numbers $n_i$ ($i=0,1,2,\ldots,M$) as
\begin{eqnarray}
Z=
\sum_{A=1}^{2^N}\exp \left (-\beta \epsilon_A \right ) 
=
\sum_{i=0}^{M} n_i \exp \left (-\beta  E_i \right ) 
=
\omega^{-\frac{M}{2}}\sum_{i=0}^{M} n_i \omega^i 
\ \ (\omega\equiv e^{-\beta}).
\label{z}
\end{eqnarray}
This means that the partition function of DREM can be 
completely determined by a set of occupation 
numbers $\{n_i \}$, in which details of the energy 
configuration $\{\epsilon_A\}$ are ignored. 
This makes it possible to assess the moments 
$\left \langle Z^n \right \rangle$ directly from 
$\{n_i \}$ without referring to the full energy configuration 
$\{\epsilon_A\}$. This significantly reduces the necessary cost for 
computing the partition function when the calculation 
is numerically performed. 

Two methods are known for generating $\{n_i\}$. 
The straightforward method is to count $n_i$ independently,
drawing the $2^N$ energy states from Eq.~(\ref{prob}). 
The system obtained by this is referred to as 
the canonical discrete random energy model (CDREM). 
Although this yields a rigorously correct realization 
of DREM that satisfies the constraint $\sum_{i=0}^{M} n_i = 2^N$, 
it takes $2^N$ steps to count $\{n_i\}$ and hence, 
is computationally difficult. In order to resolve this difficulty
in numerical experiments, 
Moukarzel and Parga  \cite{Mou1,Mou2} 
have proposed the grand canonical version of 
the discrete random energy model 
(GCDREM)\footnote{Employment of GCDREM is not essential 
for reducing the numerical cost. We have discovered a scheme 
for generating CDREM in a time scale similar to that 
for GCDREM. It is shown in appendix \ref{CDREM}.}.  
In GCDREM, the occupation numbers are independently 
determined using the Poisson distribution
\begin{eqnarray}
P(n_i) = e^{-\gamma _i}\frac{\gamma _i^{n_i}}{n_i!},
\ \ (\gamma _i = 2^N P(E_i)),
\label{Poisson}
\end{eqnarray}
where $\gamma_i$ is the average occupation number. 
The greatest advantage of GCDREM is 
that one can drastically reduce the necessary 
computational cost for generating $\{ n_i \}$ from $2^N$ to $M+1$. 
One possible drawback is that the constraint 
$\sum_{i=0}^{M} n_i = 2^N$ is only satisfied in average, 
$\left\langle\sum_{i=0}^{M} n_i\right\rangle = 2^N$, 
which implies that this method does not correspond strictly to the 
original model. 
%%%%%%% revised by YK 04/03/02
However, the RM-based calculation indicates that 
thermodynamical properties of GCDREM 
become identical to those of CDREM for 
$N \to \infty$, which is provided in appendix \ref{GCDREM}, 
and one can show that the difference rapidly vanishes as $N$ becomes 
large and is almost indistinguishable even for $N=3$, 
verified in appendix \ref{CDREM}. 
%%%%%%%
In addition, this version of DREM has another advantage 
in analytic calculations as the summation can be carried out 
independently, as is shown below.

\begin{figure}[t]
\centerline{\includegraphics[width=5cm]{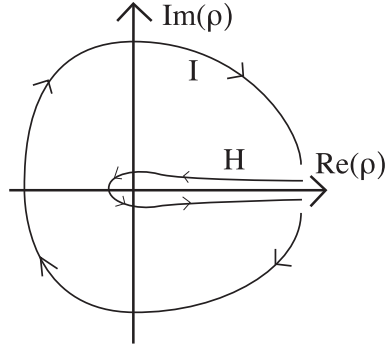}}
\caption{The integration contours.}
\label{cont}
\end{figure}

In GCDREM, the moment is expressed as
\begin{eqnarray}
\left\langle Z^n \right\rangle
&=&
\sum_{n_0=0}^{\infty}
\sum_{n_1=0}^{\infty}
\cdots
\sum_{n_M=0}^{\infty}
P(n_0)
P(n_1)
\cdots
P(n_M)
Z^n \cr
&=&
\omega^{-\frac{nM}{2}}
\lim_{\epsilon \rightarrow 0}
\sum_{n_0=0}^{\infty}
\sum_{n_1=0}^{\infty}
\cdots
\sum_{n_M=0}^{\infty}
P(n_0)
P(n_1)
\cdots
P(n_M)
\left (\sum_{i=0}^{M} n_i \omega^i+\epsilon \right )^n,
\end{eqnarray}
%%%%%% revised by YK 04/03/03.
where an infinitesimal constant $\epsilon >0$ is 
introduced in order to keep $Z$ positive
even when all the occupation numbers vanish. 
This makes it possible to employ an identity for the positive number $c$
\begin{eqnarray}
c^n
=
\frac{\int_H (-\rho)^{-n-1}
e^{-c\rho}d\rho}{\tilde \Gamma (-n)}, \ \ 
(c>0,\tilde\Gamma (n)\equiv -2i\sin n\pi \Gamma (n)),\label{cn}
\end{eqnarray}
where the integration contour is shown in Fig.~\ref{cont},
for assessing the moment as
\begin{eqnarray}
\left\langle Z^n \right\rangle
&=&
\frac{
\omega^{-\frac{nM}{2}}
}{
\tilde\Gamma (-n)
}
\lim_{\epsilon \rightarrow 0}
\int_H (-\rho)^{-n-1}e^{-(\sum_{i=0}^{M} n_i 
\omega^i+\epsilon)\rho}d\rho \cr
&=&
\frac{
\omega^{-\frac{nM}{2}}
}{
\tilde\Gamma (-n)
}
\lim_{\epsilon \rightarrow 0}
\int_H (-\rho)^{-n-1}e^{-\epsilon\rho}
\left(
\sum_{n_0=0}^{\infty}P(n_0)e^{-n_0\rho}
\right)\left(
\sum_{n_1=0}^{\infty}P(n_1)e^{-n_1\omega\rho}
\right) \cr 
&&\cdots\left(
\sum_{n_M=0}^{\infty}P(n_M)e^{-n_M\omega^M\rho}
\right)
d\rho.
\end{eqnarray}
%%%%%% 
It is worth noting that the summation 
in this expression can be independently carried out as
\begin{eqnarray}
\sum_{n_i=0}^{\infty}P(n_i)e^{-n_i\omega^i\rho}
=
e^{-\gamma_i}\sum_{n_i=0}^{\infty}\frac{1}{n_i!}
(\gamma_ie^{-\omega^i\rho})^{n_i}
=
\exp{[{-(1-e^{-\omega^i\rho})\gamma_i}]}.
\end{eqnarray}
Therefore, the moment can be summarized as 
\begin{eqnarray}
\left\langle Z^n \right\rangle
&=&
\frac{
\omega^{-\frac{nM}{2}}
}{
\tilde\Gamma (-n)
}
\lim_{\epsilon \rightarrow 0}
\int_H(-\rho)^{-n-1}\exp{[-\epsilon \rho-\sum_{i=0}^{M}
(1-e^{-\omega^i \rho})\gamma _i]}d\rho.
\end{eqnarray}
Since this is convergent for $\Re(n)>0$, 
the following expression gives the analytic 
continuation of the moment to the right half complex 
%plain revised by YK 04/03/01
plane
of $n$,
\begin{eqnarray}
\left\langle Z^n \right\rangle
&=&
\frac{
\omega^{-\frac{nM}{2}}
}{
\tilde\Gamma (-n)
}
\int_H(-\rho)^{-n-1}\exp{[-\sum_{i=0}^{M}
(1-e^{-\omega^i \rho})\gamma _i]}d\rho.
\label{momcont}
\end{eqnarray}

\subsection{Thermodynamic limit}
Using Eq.~(\ref{momcont}), one can analytically 
examine the behavior of the moment in the thermodynamic limit 
$N,M \to \infty$, keeping $\alpha=M/N$ finite. 
For this, we first convert the contour integration 
in the expression to an integration on the real axis for $p-1<\Re(n)<p$, 
where $p$ is an arbitrary natural number. 
Further, we define a function
\begin{eqnarray}
f(\rho)
\equiv
\exp{[-\sum_{i=0}^{M}(1-e^{-\omega^i \rho})\gamma _i]}
\equiv
\sum_{j=0}^{\infty}
f_i\rho^i, 
\end{eqnarray}
and a series of truncated summations as
\begin{eqnarray}
f^{(p)}(\rho)
\equiv
\sum_{j=0}^{p-1}
f_i\rho^i, 
\end{eqnarray}
which satisfy the identity
\begin{eqnarray}
\int_H(-\rho)^{-n-1}f^{(p)}(\rho)d\rho
=
\int_{H+I}(-\rho)^{-n-1}f^{(p)}(\rho)d\rho
=
0,
\label{eq:identity}
\end{eqnarray}
for $p-1 < \Re(n)$, since the contribution from the 
contour $I$ vanishes.
Using this identity, the moment can be rewritten as
\begin{eqnarray}
\left\langle Z^n \right\rangle
&=&
\frac{
\omega^{-\frac{nM}{2}}
}{
\tilde\Gamma (-n)
}
\int_H(-\rho)^{-n-1}
[f(\rho)-f^{(p)}(\rho)]
d\rho.
\end{eqnarray}
Eq.~(\ref{eq:identity}) guarantees that 
the infrared divergence is removed for $\Re(n)<p$ in this expression. 
Therefore, the moment can be rewritten for $p-1<\Re(n)<p$ as
\begin{eqnarray}
\left\langle Z^n \right\rangle
&=&
\frac{
\omega^{-\frac{nM}{2}}
}{
\Gamma (-n)
}
\int_0^{\infty}\rho^{-n-1}
[f(\rho)-f^{(p)}(\rho)]
d\rho, 
\label{eq:expression_for_p}
\end{eqnarray}
where the function $\tilde\Gamma$ is replaced by 
the ordinary gamma function $\Gamma$. 

As we have a particular interest in the case of $p=1$, 
i.e. $0<\Re(n)<1$, let us focus on 
the behavior of the following expression,
\begin{eqnarray}
\left\langle Z^n\right\rangle
&=&
\frac{
\omega^{-\frac{nM}{2}}
}{
\Gamma (-n)
}
\int_0^{\infty}\rho^{-n-1}
[f(\rho)-1]
d\rho \cr 
&=&
\frac{
\omega^{-\frac{nM}{2}}
}{
\Gamma (-n)
}
\int_0^{\infty}\rho^{-n-1}
[\exp{\{-\sum_{i=0}^{M}(1-e^{-\omega^i \rho})\gamma _i\}}-1]
d\rho. 
\end{eqnarray}
To examine the behavior of the thermodynamic limit, 
it is convenient to introduce new variables
\begin{eqnarray}
x\equiv\frac{i}{N\alpha},\ \ y\equiv\frac{1}{N \alpha \beta }\ln \rho, 
\end{eqnarray}
instead of $i$ and $\rho$, 
yielding the following expression for the moment
\begin{eqnarray}
\left \langle Z^n \right \rangle 
&=&
\frac{
\omega^{-\frac{nM}{2}}
}{
\Gamma (-n)
}
\int_{-\infty}^{\infty}e^{N{\cal G}(y)}dy,
\label{znint}
\end{eqnarray}
where
\begin{eqnarray}
\left\{
\begin{array}{ll}
{\cal G}(y)
&=
-n\alpha \beta y +\ds\frac{1}{N}\log (1-e^{-{\cal F}(y)}),\\\\
{\cal F}(y) 
&=
%%%\ds\int_0^1 e^{N{\cal H}(x)}dx,\\\\ revised by YK 04/03/01
\ds\int_0^1 e^{N{\cal H}(x,y)}dx,\\\\
{\cal H}(x,y)
&=
(1-\alpha)\log 2
+
\alpha H(x)
+
\ds\frac{1}{N}\log (1-e^{-e^{N\alpha \beta(y-x)}}), 
\end{array}
\right.
\label{eq:details}
\end{eqnarray}
and
\begin{eqnarray}
H(x)=-x \log x-(1-x) \log (1-x). 
\label{Hx}
\end{eqnarray}
Here, we have replaced the summation with an integration, 
which can be verified when both $M$ and $N$ are sufficiently large.

\begin{figure}[t] 
\centerline{\includegraphics[width=5cm]{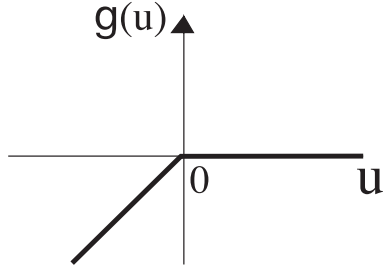}}
\caption{The shape of $g(u)$ for $N\rightarrow \infty$. 
This is not analytic at $u=0$.  }
\label{ufunc}
\end{figure}

For further analysis, the identity 
\begin{eqnarray}
%%%% revised by YK 04/03/01
%g(u)
%\equiv
%1-e^{-e^{Nu}}
%&\rightarrow &
%\left\{
%\begin{array}{c}
%1\ \ \ (u>0)\\\\
%e^{Nu}\ \ \ (u<0)\\
%\end{array}
%\right. \cr 
%\nonumber\\
%&=&
%\theta (u)+e^{Nu}\theta (-u)\ \ \ (N\rightarrow \infty), 
g(u)\equiv \frac{1}{N}\log \left (1-e^{-e^{Nu}}\right )
&\to& 
\left \{
\begin{array}{cl}
0 & (u \ge 0)\cr
u & (u < 0) 
\end{array}
\right . \cr
&=& u \theta(-u)
\label{eq:g_u}
\end{eqnarray}
which holds for large $N$, may be useful. 
The shape of this function is displayed in Fig.~\ref{ufunc}.  
As this becomes singular at $u=0$, 
the phase transition may occur in 
$n$ space for $N\rightarrow\infty$
We examine this below. 

\begin{figure}[t]
\centerline{\includegraphics[width=8cm]{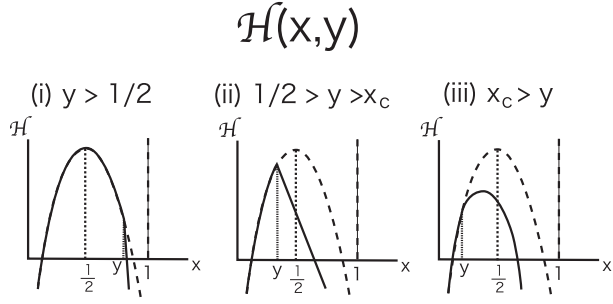}}
\caption{The schematic shape of ${\cal H}(x,y)$ 
for $N\rightarrow \infty$. It depends on the value of $y$.  }
\label{h}
\end{figure}

\begin{figure}[t]
\centerline{\includegraphics[width=5cm]{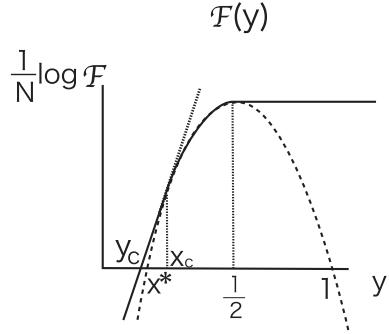}}
\caption{The schematic shape of ${\cal F}(y)$ 
for $N\rightarrow \infty$ obtained from 
the maximal value of ${\cal H}(x,y)$. }
\label{f}
\end{figure}

Eq.~(\ref{eq:g_u}) means that the function ${\cal H}$ 
can be expressed as 
\begin{eqnarray}
{\cal H}(x,y)=(1-\alpha)\log{2}+\alpha H(x)
+\alpha\beta(y-x)\theta(y-x)
\end{eqnarray}
in the thermodynamic limit.
As a function of $x$, this exhibits three behaviors 
depending on a given value of $y$, as shown in Fig.\ref{h}.  
Employing the saddle point method, 
the maximum of ${\cal H}(x,y)$ given $y$ provides ${\cal F}(y)$ 
for large $N$, which yields
\begin{eqnarray}
\lim_{N\rightarrow\infty}
\frac{1}{N}\log{{\cal F}}(y)=
\left\{
\begin{array}{c}
\ds\frac{1}{2}\ \ \left (y>\ds\frac{1}{2} \right ),\\\\
(1-\alpha)\log{2}+\alpha H(x)\ \ \left (x_c<y<\ds\frac{1}{2} \right ),\\\\
(1-\alpha)\log{2}+\alpha {\tilde H}(\beta) + \alpha \beta y\ \ 
\left (y<x_c \right ),
\end{array}
\right. 
%q(H^\prime(x_c)\equiv\beta)\label{xcdef}
\end{eqnarray}
%%%%%%%%%%
%%%%%%%%%%
%%%%%%%%%% What is q((H^\prime(x_c)\equiv\beta)?
%%%%%%%%%%
%%%%%%%%%%
where $x_c$ is a solution of 
\begin{equation}
H^\prime(x_c)=\beta, 
\label{xcdef}
\end{equation}
the behavior of which is shown in Fig.~\ref{f}.  
Here, the function ${\tilde H}$ is the Legendre transformation 
of the function ${H}$,
\begin{eqnarray}
{\tilde H}(\beta) =
\log{\left (2\cosh{\frac{\beta}{2}} \right )}
-\frac{\beta}{2}. 
\end{eqnarray}
The function ${\cal G}$ becomes 
\begin{eqnarray}
{\cal G}(y)
=
-n\alpha \beta y +\frac{1}{N}\log {\cal F}(y)\theta (-{\cal F}(y))
\label{eq:G_y}
\end{eqnarray}
for large $N$, which directly controls the behavior of 
the moment in Eq.~(\ref{znint}). 
The behavior strongly depends on the relation between $x_c$ and $x^{*}
=\left (1-\tanh\frac{\beta_c}{2} \right )/2$
which satisfies the condition 
\begin{eqnarray}
(1-\alpha)\log{2}+\alpha H(x^{*})=0.\label{x*}
\end{eqnarray}

%%%%%%%%
%%%%%%%%

\begin{itemize}
\item (A) $x^{*}<x_c$\\
Since $x_c$ and $x^{*}$ are defined in Eq.~(\ref{xcdef}) 
and Eq.~(\ref{x*}), respectively, 
the condition $x^{*}<x_c$ can be written as
\begin{eqnarray}
(1-\alpha)\log{2}+ \alpha H(x_c)>0.
\end{eqnarray}
This is satisfied for $\beta<\beta_c$, i.e. 
the high temperature phase.  
Notice that for $\alpha\leq 1$, as Eq.~(\ref{cdef})
does not have a positive solution, this is always satisfied
independently of $\beta$. 
Then, Eq.~(\ref{eq:G_y})
can be represented as 
\begin{eqnarray}
{\cal G}(y)
=
-n\alpha \beta y +[(1-\alpha)\log{2}+\alpha 
{\tilde H}(\beta) + \alpha \beta y]\theta (y_c-y),
\end{eqnarray}
which is shown in Fig.~\ref{g}.  
Here, $y_c$ is defined by ${\cal F}(y_c)=0$ (Fig.~\ref{f}), 
which yields 
\begin{eqnarray}
y_c =
-\frac{1}{\alpha\beta}[(1-\alpha)\log{2}+\alpha {\tilde H}(\beta)].
\end{eqnarray}
The moment is then calculated as
 \begin{eqnarray}
\left\langle Z^n\right\rangle
&=&
-N\alpha\beta
\frac{
\omega^{-\frac{nM}{2}}
}{
\Gamma (-n)
}
\left[
\int_{-\infty}^{y_c}
\exp{(1-\alpha)\log{2}+\alpha {\tilde H}(\beta) + \alpha \beta (1-n)y}dy\nonumber\right. \cr 
&&\hspace*{3cm}\left.
+\int^{\infty}_{y_c}\exp{(-n\alpha\beta y)}dy
\right]\\
&\sim&
e^{\alpha\beta nN(\frac{1}{2}-y_c)}
\left[\frac{1}{\Gamma (1-n)}+\frac{n}{\Gamma (2-n)}
\right].
\end{eqnarray}
Therefore, the asymptotic behavior of the moment is
\begin{eqnarray}
\lim_{N\rightarrow\infty}
\frac{1}{N}\log{\left\langle Z^n\right\rangle}
&=&
\alpha\beta nN \left (\frac{1}{2}-y_c \right ) \cr 
&=&
n \left (\log{2}+\alpha \log{\cosh{\frac{\beta}{2}}}  \right )
\label{rs1_ours}
\end{eqnarray}
in this phase.  
This is consistent with RS1, as can be seen from Eq.~(\ref{rs1}).

%%%%%%%
\begin{figure}[t]
\centerline{\includegraphics[width=10cm]{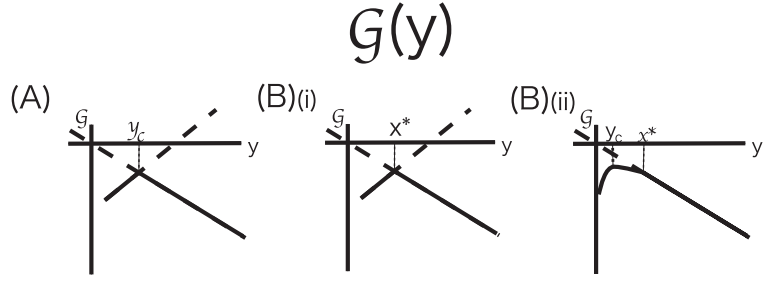}}
\caption{The schematic figure of ${\cal G}(y)$.  
The maximal value determines the behavior of 
$\left \langle Z^n \right \rangle$. }
\label{g}
\end{figure}
%%%%%%%

\item (B) $x^{*}>x_c$\\
The condition $x^{*}>x_c$ is satisfied for 
$\beta>\beta_c$, which may correspond to the low temperature phase.
The function ${\cal G}(y)$ has the form
\begin{eqnarray}
{\cal G}(y) =
-n\alpha \beta y +[(1-\alpha)\log{2}+\alpha H(y)]\theta (x^{*}-y), 
\end{eqnarray}
the behavior of which is further classified into two 
cases depending on the replica number $n$. 

\begin{itemize}
\item (i) $H^\prime(x^{*})>n\beta$\\
Since $x^{*}$ is located in $x_c<x^{*}< \frac{1}{2}$, 
there exists a critical replica number $0<n_c<1$, 
which is characterized by 
\begin{eqnarray}
%n_c = \frac{1}{\beta} H^\prime(x^{*}). 
n_c  \equiv \frac{1}{\beta} H^\prime(x^{*})=\frac{\beta_c}{\beta}. 
\end{eqnarray}
The condition $H^\prime(x^{*})>n\beta$ is satisfied for $n<n_c$.  
The function ${\cal G}(y)$, shown in Fig.~\ref{g}, 
is maximized at $y=x^{*}$. Therefore, 
the moment can be expressed as  
\begin{eqnarray}
\lim_{N\rightarrow\infty}
\frac{1}{N}\log{\left\langle  Z^n\right\rangle} &=&
n\alpha\beta\left (\frac{1}{2}-x^{*} \right ) \cr
&=& \frac{n \alpha \beta}{2} \tanh\frac{\beta_c}{2}. 
\label{1rsb_ours}
\end{eqnarray}
This behavior is identical to that of
1RSB predicted by RM, as can be seen in Eq.~(\ref{1rsb}).

\item (ii) $H^\prime(x^{*})<n\beta$\\
The condition $H^\prime(x^{*})<n\beta$ is satisfied for $n>n_c$.
The function ${\cal G}(y)$ is maximized
not at $y=x^{*}$ but at $y=y_c$.  
Therefore, the moment can be asymptotically expressed as 
\begin{eqnarray}
\lim_{N\rightarrow\infty}
\frac{1}{N}\log{\left\langle Z^n\right\rangle}
&=&
%%%%%%%\log{2}+\log{\cosh{\frac{n\beta}{2}}}.
\log{2}+ \alpha \log{\cosh{\frac{n\beta}{2}}}.
\label{rs2_ours}
\end{eqnarray}
This coincides with the behavior of RS2 obtained by RM, 
as can be seen in Eq.~(\ref{rs2}).
\end{itemize}
\end{itemize}

\begin{figure}[t]
\centerline{\includegraphics[width=12cm]{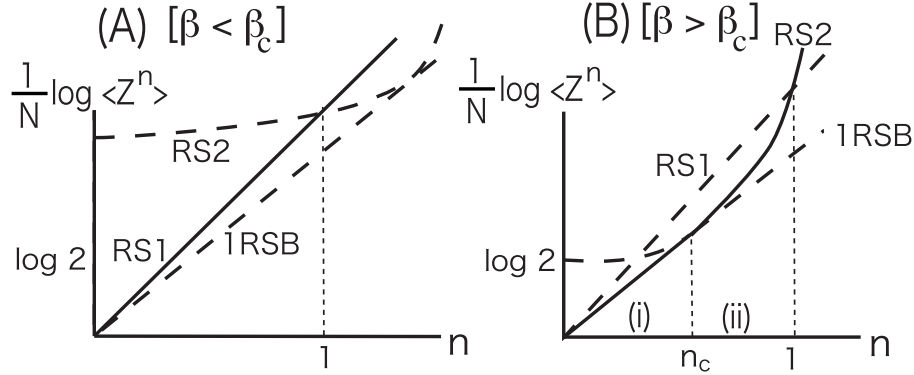}}
\caption{The behavior of the moment in the thermodynamic limit
obtained from the exact expression (full lines).  
In the high temperature phase 
$\beta<\beta_c$, the behavior of the moment is simple and 
corresponds to the result obtained from RS1.  In the low temperature
phase $\beta>\beta_c$, the behavior of the moment has interesting
properties.  The moment corresponds to that obtained from RS2 for
$n>n_c \equiv \beta/\beta_c$ and corresponds to 
that obtained from 1RSB for $n<n_c$. A phase transition occurs at $n=n_c$.}
\label{phase}
\end{figure}

The results are summarized in Fig.~\ref{phase}.  
In the high temperature phase $\beta<\beta_c$, 
the behavior of the moment is simple, being expressed 
by RS1 of RM. In the low temperature phase $\beta>\beta_c$, 
the behavior of the moment has two possibilities 
depending on $n$. More specifically, in the limit $N \to \infty$, 
the moment approaches RS2 for $n > n_c$, whereas 1RSB represents 
the correct behavior for $n<n_c$. This means that there exists 
a phase transition in the space of replica number at $n=n_c$. 
In conclusion, these are consistent with the known results 
obtained by RM at the limit $n \to 0$ \cite{REM,Mou1,Mou2}. 

\subsection{Numerical validation}
Eq.~(\ref{momcont}) is formulated as a two dimensional summation 
with respect to $\rho \in {\bf C}$ and $i=0,1,\ldots,M$, 
which is numerically tractable. This means that one can utilize 
Eq.~(\ref{momcont}) or Eq.~(\ref{znint}) to numerically examine
the behavior of DREM for a finite system size $N$, 
and how fast the results obtained for $N \to \infty$ 
become relevant as $N$ grows large. 

\begin{figure}[t]
\centerline{\includegraphics[width=8cm]{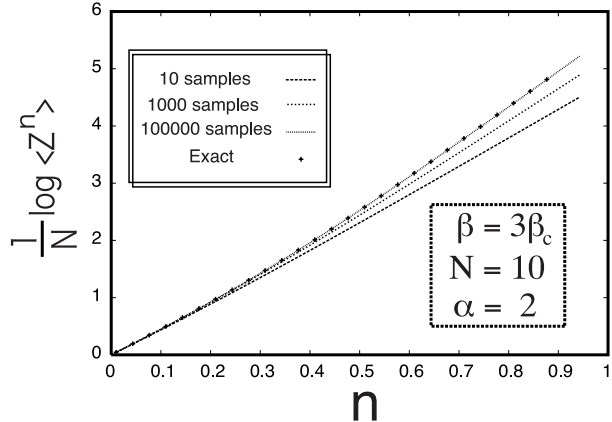}}
\caption{The logarithm of $\left \langle Z^n \right \rangle$ 
calculated from Eq.~(\ref{znint}) and 
data obtained from numerical experiments of Eq.~(\ref{z}) 
using 10, 1000 and 100000 samples. The system size is $N=10$. 
The points indicated as ``Exact"
are the results obtained from our expression Eq.~(\ref{znint}).}
\label{sam}
\end{figure}

Fig.~\ref{sam} shows the logarithm of 
$\left \langle Z^n \right \rangle$ calculated by 
Eq.~(\ref{znint}) and $\left \langle Z^n \right \rangle$ 
numerically evaluated from 10, 1000 and 100000 
experiments, with $N=10$. 
One can see that the data from the numerical experiments 
converge to the results of Eq.~(\ref{znint}). 
This verifies that our expression accurately 
provides the moment even for a finite system size. 

\begin{figure}
\centerline{\includegraphics[width=6cm]{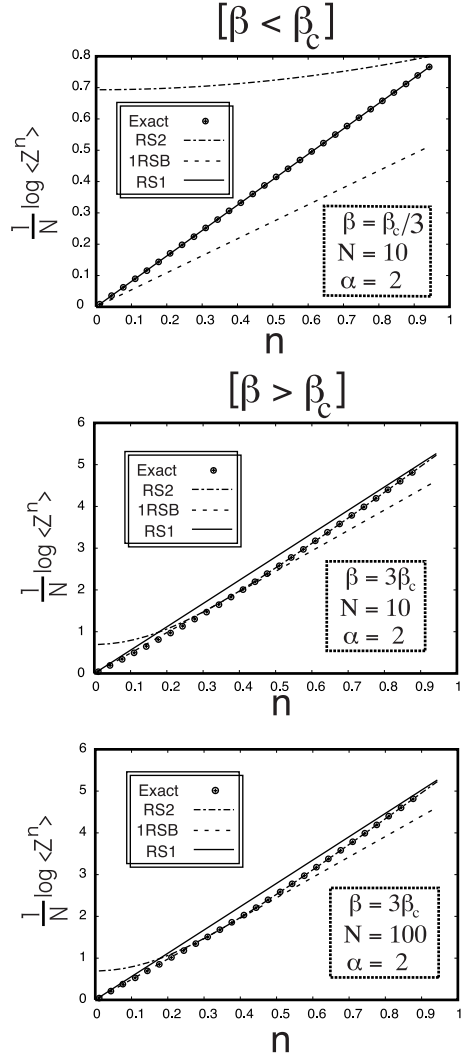}}
\caption{
The logarithm of $\left \langle Z^n \right \rangle$  
obtained from our expression and RM.  The points indicated as ``Exact"
are the results obtained from our expression Eq.~(\ref{znint}).  
The lines indicated ``RS1", 
``RS2", and ``1RSB" are the results obtained from RM, 
using Eq.~(\ref{rs1}), Eq.~(\ref{rs2}), and Eq.~(\ref{1rsb}), respectively.
At high temperatures $\beta=\beta_c/3<\beta_c$, our result is consistent with 
RS1. The difference is very small even for $N=10$.
At low temperatures $\beta=3\beta_c > \beta$, our result fits 
RS2 for larger $n>n_c=\beta_c/\beta=1/3$ while 
1RSB provides good consistency for smaller $n<n_c$. 
There is a slight difference between the results from 
our expression and that from 1RSB at $N=10$. 
The difference, however, becomes imperceptible for $N=100$.}
\label{com}
\end{figure}

We next compare the results from our expression and 
RM in Fig.~\ref{com}.  
At high temperatures, our result is consistent with RS1 as expected. 
The difference is negligible for all of the range $0<n<1$ even at $N=10$.
At low temperatures, our result fits RS2 for larger $n>n_c$, 
while 1RSB exhibits 
excellent consistency for smaller $n<n_c$. 
There is a slight difference between our expression and 1RSB for $N=10$.  
The difference, however, becomes indistinguishable for $N=100$. 
This strongly indicates that there occurs a phase transition 
between RS2 and 1RSB at $n=n_c$ in the limit $N \to \infty$.

\section{Origin of 1RSB: Extreme value statistics}\label{thermo}
The preceding two sections indicate that 
DREM exhibits a phase transition with respect to the replica 
number $n$ at a certain critical point $n_c \in [0,1]$ 
when the temperature is sufficiently low. In this section, 
we discuss how this transition can be understood 
in the framework of RM. A formalism previously introduced 
for examining the domain size distribution 
of multi-layer perceptrons is useful for this purpose \cite{Monasson}. 

Since the partition function of DREM typically scales 
exponentially with respect to $N$, we first express 
this dependence as 
\begin{eqnarray}
Z \sim \exp \left [-N\alpha \beta \left (y -\frac{1}{2} \right )\right ], 
\label{eq:free_energy}
\end{eqnarray}
where $y-\frac{1}{2}$ represents the free energy 
normalized by the scale of the energy amplitude $M$
for a given realization $\{\epsilon_A\}$ ($A=1,2,\ldots,2^N)$ 
(or $\{n_i \}$ ($i=0,1,\ldots,M$)). 
Clearly, $y$ is a random variable. Let us assume that 
the probability distribution of $y$, ${\cal P}(y)$, scales as
\begin{eqnarray}
{\cal P}(y) \sim \exp \left [ - N c(y) \right ], 
\label{eq:free_probability}
\end{eqnarray}
where $c(y) \sim O(1)$ for large $N$. 
Notice that an inequality 
\begin{eqnarray}
c(y) \ge 0, 
\label{eq:c_inequality}
\end{eqnarray}
must hold to make 
${\cal P}(y)$ satisfy the normalization 
condition $\int dy {\cal P}(y)=1$ for $N \to \infty$. 
Eq. (\ref{eq:free_probability}) 
indicates that the moment of the partition function 
can be calculated as 
\begin{eqnarray}
\left \langle Z^n \right \rangle &\equiv&
\int dy \exp \left [-Nn\alpha \beta 
\left (y -\frac{1}{2} \right )\right ]\cP(y) \cr
&\sim & \exp \left [N \mathop{\rm extr}_{y}
\left \{- n \alpha \beta \left (y-\frac{1}{2}\right )  
- c(y) \right \} \right ]. 
\label{eq:free_legendre}
\end{eqnarray}
This formula, however, may not be useful for computing 
$\left \langle Z^n \right \rangle$, as directly assessing $c(y)$ 
is rather difficult. 
Instead, it can be utilized to evaluate $c(y)$ 
from $\left \langle Z^n \right \rangle$, 
since Eq.~(\ref{eq:free_legendre})
implies that $c(y)$ can be obtained from 
\begin{eqnarray}
c(y)= \mathop{\rm extr}_{n}
\left \{ -n \alpha \beta \left (y-\frac{1}{2}\right ) 
-\frac{1}{N} \ln \left \langle Z^n \right \rangle \right \}, 
\label{eq:c_g}
\end{eqnarray}
where $(1/N) \ln \left \langle Z^n \right \rangle$
can be computed by RM for any real number $n$. 
This, in conjunction with the normalization constraint 
in Eq.~(\ref{eq:c_inequality}), offers a useful clue to 
identify the origin of the phase transition 
with the aid of RM. 

For $\beta > \beta_c$, RS2 provides the dominant solution 
for natural numbers $n=1,2,\ldots$ in the thermodynamic limt. 
Carlson's theorem implies that this should offer a unique 
analytic continuation if no phase transition occurs. 
Therefore, let us first insert this solution, from Eq.~(\ref{rs2}), 
into Eq.~(\ref{eq:c_g}), giving 
\begin{eqnarray}
c(y)&=&\mathop{\rm extr}_{n}
\left \{ -n \alpha \beta \left (y-\frac{1}{2} \right ) -\log 2 -
\alpha \log{ \left (\cosh{\frac{n\beta}{2} } \right ) } \right \} \cr
&=& (\alpha -1) \log 2 - 
\alpha H(y). 
\label{eq:extreme_value_stat}
\end{eqnarray}
Notice that this does not satisfy the necessary 
condition given in Eq.~(\ref{eq:c_inequality}) for $y > x^*$, and 
the critical value $y=x^*$ corresponds to 
$n=n_c=H^\prime(x^*)/\beta=\beta_c/\beta$, 
which signals the occurrence of a phase transition 
at $n=n_c$ within the framework of RM. 

The following consideration about the energy configuration 
provides a plausible scenario 
for this transition. As have already been pointed out \cite{Mezard_Extreme}, 
the partition function $Z$ in the low temperature region $\beta > \beta_c$, 
can be considered to be dominated by only the minimum 
energy $\epsilon_{\rm \tiny min}$ of 
a given energy configuration $\{\epsilon_A \}$, such as 
$Z \sim \exp \left [ -\beta \epsilon_{\rm \tiny min} 
\right ]$. This implies that 
$\cP(y)$ of the normalized free energy 
$y-\frac{1}{2}=-(\log Z)/(N\alpha\beta)$ 
is given by the distribution of $\epsilon_{\rm \tiny min}/M$ 
under the assumption that each energy 
level $\epsilon_A$ $(A=1,2, \ldots, 2^N)$ is 
independently drawn from an identical distribution, given in Eq.~(\ref{prob}), 
which has been already examined in the research 
of {\em extreme value statistics} (EVS)
\cite{extreme_value}. In the current case, the average 
occupation number of an energy level $E_i=i-\frac{1}{2}M=M 
\left (x-\frac{1}{2} \right )$, $\gamma_i=2^{N-M} 
\left ( 
\begin{array}{c}
M \cr
M/2+E_i 
\end{array}
\right )
=\exp N
\left ((1-\alpha)\log 2 + \alpha H(x) \right )$
($i=0,1,\ldots,M$ and $x=i/M \in [0,1]$), 
grows exponentially large with respect to $N$ for $x>x^*$. 
This indicates that $\cP(y)$ is very small 
for $y > x^*$ because $x>x^*$ does not provide 
$\epsilon_{\rm \tiny min}/M$ 
for a given energy configuration $\{ \epsilon_{A} \}$
except for very rare cases, as 
energy levels lower 
than $E_i=M\left (x-\frac{1}{2} \right )$ 
are included in the configuration with 
a very high probability. On the other hand, $\gamma_i$ becomes exponentially 
small for $0<x<x^*$, which means that the probability 
of having an energy level labelled by $x$ of this interval in 
the energy configuration is low. Therefore, $\cP(y)$ 
also becomes small for $0<y<x^*$. However, the functional form 
of $\cP(y)$ is not symmetric between $y>x^*$ and $0< y < x^*$. 
Actually, detailed analysis of EVS\cite{extreme_value,Mezard_Extreme} 
shows that $\cP(y)$ exhibits the asymmetric 
scalings
\begin{eqnarray}
\cP(y) \sim \left \{
\begin{array}{ll}
\exp \left [- \exp \left [ -N \left ((\alpha -1) \log 2 - 
\alpha H(y)  \right ) \right ] \right ], & (\mbox{$x^*<y$}), \cr
\exp \left [ -N \left ((\alpha -1) \log 2 - 
\alpha H(y)  \right ) \right ], & (\mbox{$0< y< x^*$}), 
\end{array}
\right .
\label{eq:P_y}
\end{eqnarray}
for large $N$, which provide a singularity at $y=x^*$. 
%%%%%%%%
%%%%%%%% YK -added 
%%%%%%%%
It may be worth noting that 
$1 - \exp \left [- {\cal F}(y) \right ]$, which comes out in 
%%%%% revised by YK 04/03/04
Eq.~(\ref{znint}) after inserting Eq.~(\ref{eq:details})
in the analysis presented section \ref{exact}, 
corresponds to the cumulative distribution 
$\int_{-\infty}^y d\tilde{y} {\cal P}(\tilde{y})$. 
%%%%%
Eq.~(\ref{eq:P_y}) implies that the extremum 
point in Eq.~(\ref{eq:free_legendre})
for large $n(>n_c)$ is in $0<y < x^*$, which leads to 
RS2 and consistently provides the exponent of 
Eq.~(\ref{eq:extreme_value_stat}).
On the other hand, the constant $y=x^*$ offers 
the extremum for all small $n(<n_c)$, which corresponds 
to 1RSB in the framework of RM. 
The intuitive implication of this is that $\left \langle Z^n \right \rangle$ 
is dominated by an atypically low minimum energy generated 
with a small probability for $n > n_c$, whereas
the typical minimum energy $\epsilon_{\rm \tiny min}=
M \left (x^*-\frac{1}{2} \right )$ 
provides the major contribution to $\left \langle Z^n \right \rangle$
for $n < n_c$. Thus, the origin of the phase transition can 
be attributed to the singularity of the free energy distribution $\cP(y)$, 
exhibited in the case of low temperatures $\beta > \beta_c$. 

Our analysis also illustrates that Carlson's theorem is 
not necessarily useful for validating RM because 
%%%%%% Corrected by YK 04/03/19
an analytic continuation given for finite $N$ 
%%%%%% 
can exhibit a singularity in the limit of 
$N \to \infty$ and, therefore, taking the 
limit $N \to \infty$ prior to $n \to 0$
for determining the continuation 
on the basis of expressions for $n=1,2,\ldots$, 
which is usually performed in RM, 
sometimes yields a wrong answer for $n<1$. 
However, in the current system, 
this drawback can be resolved by taking a constraint 
for the free energy distribution 
(i.e.~Eq.~(\ref{eq:c_inequality}))
into account, which leads to the conventional 1RSB. 
Although the importance of the notion of 
EVS in RM has been already addressed 
at the limit $n \to 0$ \cite{Mezard_Extreme}, 
to our knowledge, the current analysis is the first work 
that directly clarifies how the property of 
EVS relates to a corruption of the analytic continuation 
in RM, expressed as a phase transition 
with respect to the replica number $n$. 

\section{Summary}\label{summary}
In summary, we have offered an exact expression, Eq.~(\ref{momcont}), 
of the moment of the partition function $\left\langle Z^n \right\rangle$ 
for GCDREM, without employing the replica trick, which is valid for 
any arbitrary system size $N$ and complex number $n\ (\Re(n)>0)$. 
Simplifying the expression for $0<n<1$, 
we have shown that a phase transition with respect 
to the number of replicas $n$ occurs at a certain 
critical number $n_c \in [0,1]$ in the thermodynamic 
limit $N \to \infty$ when the temperature $\beta^{-1}$ is sufficiently 
low, if the ratio $\alpha=M/N$ is greater than unity. 
This implies that Carlson's theorem, which guarantees the uniqueness 
of the analytic continuation from the expressions 
for $n=1,2,\ldots$ to those for $n \in {\bf C}$, 
is not necessarily useful for validating 
the replica method, because taking the thermodynamic 
limit $N \to \infty$ prior to the continuation 
fails to derive a correct result when the analyticity 
is broken for $n<1$ in the case of infinite $N$. 
However, it has also been shown that this drawback can be overcome
by taking the statistical property of the minimum energy level into 
account, clarifying how 1RSB originates from a replica 
symmetric solution, which has been conventionally discarded, 
at the critical replica number $n_c$. 
We hope that the results obtained here offer a useful insight 
to unlock the remaining mysteries of RM.  

Since Eq.~(\ref{momcont}) is valid 
throughout the right half complex 
plane $\Re(n) > 0$ for any finite $N$, 
one can directly observe how the singularities of 
$\left\langle Z^n \right\rangle$ approach the real axis of $n$ 
as $N \to \infty$\cite{van_Hemmen}. This is sometimes referred to as 
Lee-Yang's scenario of the phase transition\cite{Lee_Yang}. 
Analysis along this direction is currently under way\cite{Ogu}.

\section*{Acknowledgement}
Support from Grant-in-Aids of MEXT, Japan, No.~14084206 (YK) is 
acknowledged.

\appendix
\section{Link of DREM to error-correcting codes}
\label{ECC}
As was shown in Ref. \citen{Sourlas}, certain types of spin glass models 
can be related to error-correcting codes. 
We here show that a known evaluation scheme of the error 
correcting ability of a random code ensemble can 
be reduced to a calculation of the general moment of 
the partition function with respect to DREM\cite{Gallager,JPA_review}.
To the knowledge of authors, a remark on this relationship has already 
been made in Ref. \citen{Iba_note}. 

In a general scenario of error correcting codes, 
an $N$-dimensional binary vector $\bbs=(s_1,s_2,\ldots,s_N)$ 
($s_i=0,1$, $i=1,2,\ldots,N$) is encoded to a codeword
$\bt=(t_1,t_2,\ldots,t_M)$ ($t_i=0,1$, $i=1,2,\ldots,M$) 
of an $M(>N)$-dimensional binary vector and transmitted via a noisy channel. 
We here concentrate on a binary symmetric channel (BSC) in which 
each component of the codeword $t_i$ is independently 
flipped to the opposite letter in the alphabet ($0$ or $1$) 
with a probability $0<p<1/2$.
This implies that the vector $\bbr=\bt+\bn$ (mod 2) is received 
at the other terminal of the channel, where $\bn$ is the noise vector,  
the component of which becomes $1$ independently with probability 
$p$ and $0$ otherwise. However, as the codeword is represented somewhat 
redundantly, decoding $\bbr$ provides the correct message $\bbs$ 
with a high probability for sufficiently small $p$. 

It is known that the performance of correctly retrieving
$\bbs$ from $\bbr$ becomes good when a code ${\cal C}$
(i.e., an invertible map $\cal C:\bbs \leftrightarrow \bt$) 
is randomly constructed \cite{Shannon,Gallager}, providing 
a code ensemble. Let us evaluate the average 
probability of failing to correctly retrieve $\bbs$ in order to 
characterize the potential error correcting ability of the ensemble. 
We focus on the maximum likelihood (ML) decoding, 
which selects a codeword closest to the received vector 
$\bbr$ and returns a message vector corresponding to 
the codeword as an estimate $\hat{\bbs}$ of $\bbs$. 
ML decoding minimizes the decoding error probability $P_E({\cal C})$
when codeword vectors  $\bbt$ are equally generated at 
the transmission terminal \cite{Iba_JPA}. 

Notice that the message $\bbs$ can be correctly 
identified if its codeword $\bbt$ is provided, since
the code is constructed as an invertible map. 
Therefore, for evaluating $P_E(\cal C)$, 
it is convenient to introduce an indicator function 
$\Delta_{\rm \tiny ML}(\bbt,\bbr|{\cal  C})$ that returns 
$1$ if $\bbr$ is not correctly decoded to the correct 
codeword $\bbt$ and $0$ otherwise, 
which means the decoding error probability can be computed by
\begin{eqnarray}
P_E({\cal C})=\sum_{\bbt,\bbr}P(\bbt|{\cal C})P(\bbr|\bbt)
\Delta_{\rm \tiny ML}(\bbt,\bbr|{\cal  C}), 
\label{eq:block_error}
\end{eqnarray} 
and, hence, its average over the code ensemble 
can be represented as 
\begin{eqnarray}
\left \langle P_E({\cal C})
\right \rangle_{\cal C}=
\sum_{{\cal C}}P( {\cal C} )\sum_{\bbt \in {\cal C},\bbr}
P(\bbt|{\cal C})
P(\bbr|\bbt)\Delta_{\rm \tiny ML}(\bbt,\bbr|{\cal  C}), 
\label{eq:average_block_error}
\end{eqnarray} 
where $P(\bbt|{\cal C})$ is the probability 
that the codeword vector $\bbt$ is generated given ${\cal C}$
and $P(\bbr | \bbt)$ is the conditional probability 
that $\bbr$ is received when $\bbt$ is transmitted.  
$P( {\cal C} )$ is the probability that a code ${\cal C}$ 
is generated. $\sum_{\bbt \in {\cal C}}$ 
denotes a summation over $2^N$ codeword vectors given ${\cal C}$. 

We assume that the code ${\cal C}$ is designed by the 
source coding technique such that $\bbt$ is uniformly 
generated as $P(\bbt|{\cal C})=1/2^N$\cite{Gallager}. 
In addition, for BSC, the conditional probability 
can be represented as 
%%%%%% revised by YK 04/03/04
\begin{eqnarray}
P(\bbr|\bbt)=
\frac{\exp\left [-F \sum_{i=1}^M \left (\frac{1}{2}-\delta_{r_i,t_i}
\right ) \right ]}
{\left (2 \cosh \frac{F}{2}\right )^M }, 
\label{eq:conditional_BSC}
\end{eqnarray}
where $\delta_{x,y}$ returns $1$ if $x=y$ and $0$, otherwise
and $F=\log\left [(1-p)/p \right ]$. 

Unfortunately, expressing $\Delta_{\rm \tiny ML}(\bbt,\bbr|{\cal  C})$
in a rigorously treatable form is difficult. However, 
Gallager's inequality
\begin{eqnarray}
\Delta_{\rm \tiny ML}(\bbt,\bbr|{\cal  C}) 
\le 
\left (
\sum_{\bbt^\prime \in {\cal C} \backslash \bbt} 
\left (\frac{P(\bbr|\bbt^\prime )}{
P(\bbr|\bbt)} \right )^{\frac{1}{{1+n}}} \right )^n,
\label{eq:gallager1}
\end{eqnarray}
which holds for arbitrary $n \ge 0$,  
offers a good upperbound for this \cite{Gallager,JPA_review}. 
Here $\sum_{\bbt^\prime \in {\cal C} \backslash \bbt} $ 
denotes a summation over 
%$2^{N-1}$ revised by YK 04/03/01
$2^{N}-1$ 
codeword vectors $\bbt^\prime$ of ${\cal C}$ excluding the possibility 
of the correct codeword $\bbt$. 
Inserting this into Eq. (\ref{eq:average_block_error}) provides
\begin{eqnarray}
\left \langle P_E({\cal C})
\right \rangle_{\cal C} 
&\le &
\sum_{{\cal C}}P( {\cal C} ) 
\sum_{\bbt,\bbr} P(\bbt|{\cal C})
P^{\frac{1}{1+n}}(\bbr|\bbt)
\left (\sum_{\bbt^\prime \in {\cal C}\backslash \bbt}
P^{\frac{1}{1+n}}(\bbr|\bbt) \right )^n \cr
&=& \sum_{{\cal C}}P( {\cal C} ) 
\sum_{\bbr} P^{\frac{1}{1+n}}(\bbr|\bzero) 
\left (\sum_{\bbt^\prime \in {\cal C} \backslash \bzero}
P^{\frac{1}{1+n}}(\bbr|\bbt^\prime) \right )^n, \cr
&=&\sum_{r_i=\pm 1} 
\frac{
e^{\frac{F}{1+n}\sum_{i=1}^M \left (\frac{1}{2}-r_i \right ) }}
{\left (2\cosh \frac{F}{2} \right )^M}
\sum_{{\cal C}} P({\cal C}) 
\left ( \sum_{\bt^\prime \in {\cal C} \backslash \bzero} 
%\exp\left [
e^{
-\frac{F}{1+n} \sum_{i=1}^M 
\left (\frac{1}{2}-\delta_{r_i,t^\prime_i} \right ) %\right ]
}
\right )^n, 
\label{eq:gallager2}
\end{eqnarray}
where we have performed the gauge transformation 
$\bbt \to \bzero$, $\bbt^\prime - \bbt \to \bbt^\prime$ and 
$\bbr-\bbt \to \bbr$, assuming any code ${\cal C}$ 
contains the zero codeword $\bbt=\bzero$ \cite{Gallager}. 
Minimizing the final expression with respect to $n \ge 0$, 
we can obtaine the tightest bound of the average decoding error
probability. 

We here address that $E(\bbt)=\sum_{i=1}^M \left 
(\frac{1}{2}-\delta_{r_i,t_i} \right )$ 
in Eq.~(\ref{eq:gallager2}) obeys Eq.~(\ref{prob})
independetly of $\bbr$ when each codeword $\bbt$ is equally generated 
in the ensemble such that each component 
is independently selected from an identical unbiased distribution
$P(t_i=1)=P(t_i=0)=1/2$, 
%%%%%% revised by YK 04/03/04
which holds for the random code ensemble\cite{Shannon,Gallager}, 
although non-trivial correlations of `energy' $E(\bbt)$ 
between `states' $\bbt$ arise making the energy distribution 
different from Eq.~(\ref{prob}) in the case of practical linear codes. 
%%%%%%
Therefore, for the current random code ensemble, 
Eq. (\ref{eq:gallager2}) can 
be simplified as
\begin{eqnarray}
\left \langle P_E({\cal C})
\right \rangle_{\cal C}  
\le 
\left (\frac{\cosh \frac{F}{2(1+n)}}{  \cosh \frac{F}{2}  }
\right )^M
\left \langle 
\left (
\sum_{A=1}^{2^N-1} \exp \left [-\frac{F}{1+n} \epsilon_A \right ]
\right )^n \right \rangle,  
\label{eq:bound_DREM}
\end{eqnarray}
where $\left \langle \cdots \right \rangle $ denotes 
the configutational average with respect to the `energy 
states' $\epsilon_A$ ($A=1,2,\ldots,2^N -1$)
following Eq.~(\ref{prob}). 
Regarding $F/(1+n)$ as the inverse temperature, 
this means that an assessment of the average decoding error 
probability can be linked to calculating 
the general moment of the `partition function' $Z=\sum_{A=1}^{2^N-1} 
\exp \left [-\frac{F}{1+n} \epsilon_A \right ]$ of DREM. 

\section{The replica analysis of GCDREM}
\label{GCDREM}
Section \ref{exact} indicates that the generalized 
moment of partition function can be evaluated {\em without 
using RM} for GCDREM even when the system size $N$ is finite. 
However, for direct comparison to the conventional analysis, 
it may be helpful to demonstrate the conventional RM-based 
analysis as well. Therefore, we here provide a brief sketch of 
the replica calculation of GCDREM. 

From eq. (\ref{z}), we first obtain an expression 
\begin{eqnarray}
\left \langle Z^n \right \rangle 
&=&\omega^{-\frac{nM}{2}}
\sum_{i_1=0}^M\sum_{i_2=0}^M
\cdots \sum_{i_n=0}^M
\left \langle \prod_{\mu =1}^n 
n_{i_\mu } 
\right \rangle 
\omega^{\sum_{\mu =1}^n i_\mu } \cr
&=& \omega^{-\frac{nM}{2}}
\sum_{i_1=0}^M\sum_{i_2=0}^M
\cdots \sum_{i_n=0}^M
\prod_{i=1}^M 
\left (
\left \langle n_i^{\sum_{\mu=1}^n \delta_{i,i_\mu}} 
\right \rangle 
\omega^{i \sum_{\mu =1}^n \delta_{i,i_\mu } }
\right ), 
\label{replica_GCDREM}
\end{eqnarray}
for $n=1,2,\ldots$, where 
$\left \langle \cdots \right \rangle $ represents
the average over the Poission distribution, Eq. (\ref{Poisson}). 
Following the conventional recipe of RM, 
let us next evaluate the candidates of the most 
dominant contribution in the limit $N \to \infty$ under the RS1, 
RS2 and 1RSB assumptions following the conventional scheme of RM. 
Before proceeding further, it may be worth mentioning that 
the dynamical variables in GCDREM are not single states but 
energy levels $i=0,1,2,\ldots,M$, each of which is 
composed of multiple states. 
This difference makes composition of the solutions slightly different 
from that of CDREM provided in section \ref{replica}, 
although the final result is unchanged. 

\begin{itemize}
\item{RS1}\\
In RS1, all the $n$ replica levels are assumed to 
be allocated to $n$ different levels $i(=0,1,2,\ldots,M)$. 
Therefore, for a given level $i$, 
\begin{eqnarray}
\sum_{\mu=1}^n \delta_{i,i_\mu}=
\left \{
\begin{array}{ll}
1 & (\mbox{when $i$ is one of the $n$ allocated energy levels}), \cr
0 & (\mbox{otherwise}).
\end{array}
\right .
\end{eqnarray}
When $i$ is an allocated level, 
\begin{eqnarray}
\left \langle n_i^{\sum_{\mu=1}^n \delta_{i,i_\mu}} 
\right \rangle 
&=&\gamma_i \to \exp\left [N \left ( (1-\alpha)\log 2 + \alpha H (x) 
\right )\right ], 
\label{average_number_of_state} 
\end{eqnarray}
and
\begin{eqnarray}
\omega^{i \sum_{\mu =1}^n \delta_{i,i_\mu } }
&=& \exp \left [ -\beta i  \right ]
\to \exp \left [ -N \alpha \beta x \right ], 
\label{occupied_contribution}
\end{eqnarray}
hold for large $N$, 
yielding the contribution 
\begin{eqnarray}
\left \langle n_i^{\sum_{\mu=1}^n \delta_{i,i_\mu}} 
\right \rangle 
\omega^{i \sum_{\mu =1}^n \delta_{i,i_\mu } }
\to \exp \left [ N
\left ( (1-\alpha)\log 2 + \alpha( H(x)-\beta x) \right )
\right ], 
\end{eqnarray}
where $x\equiv i/M=i/(\alpha N)$. 
This is maximized at $x_c=(1-\tanh \frac{\beta}{2} )$ to
\begin{eqnarray}
\exp  N\left ( \log 2+\alpha \log \left (
\cosh\frac{\beta}{2} \right ) - \frac{\alpha\beta}{2} \right ) ,
\label{first_replica}
\end{eqnarray} 
which represents the contribution from 
a certain replica level $i_\mu(=1,2,\ldots,n)$. 
Contributions from other replicas become
smaller than Eq. (\ref{first_replica}) as
any two replica levels must be located 
at different levels under the current assumption. 
However, the difference becomes negligible 
because there exist many levels in any vicinity of 
$x=x_c$ in the limit $N\to \infty$. 
Taking the prefactor $\omega^{-\frac{nM}{2}}$ 
in front of the summation in Eq. (\ref{replica_GCDREM})
and the number of permutations 
over replica indices into account, 
this implies that the dominant contribution under 
the RS1 ansatz is evaluated as 
\begin{eqnarray}
\left \langle Z^n \right \rangle 
&= &n! \times \exp n N\left (\log 2 + \alpha \log 
\left ( \cosh \frac{\beta}{2} \right ) \right )\cr
&\sim &
\exp n N\left (\log 2 + \alpha \log 
\left ( \cosh \frac{\beta}{2} \right ) \right ), 
\label{RS1_appendix}
\end{eqnarray}
which is identical to the RS1 solution of CDREM, Eq. (\ref{rs1}), 
and equivalent to Eq. (\ref{rs1_ours}). 

\item{RS2}\\
In RS2, all the $n$ replica levels are assumed to be allocated
to a certain single level. 
Therefore, for a given level $i$, 
\begin{eqnarray}
\sum_{\mu=1}^n \delta_{i,i_\mu}=
\left \{
\begin{array}{ll}
n & (\mbox{when $i$ is the allocated energy level}), \cr
0 & (\mbox{otherwise}).
\end{array}
\right .
\end{eqnarray}
When $i$ is the allocated level, 
\begin{eqnarray}
\left \langle n_i^{\sum_{\mu=1}^n \delta_{i,i_\mu}} 
\right \rangle 
&\sim &
\left \{
\begin{array}{ll}
\gamma_i^n 
\to \exp\left [nN \left ( (1-\alpha)\log 2 + \alpha H (x) 
\right )\right ]& (x > x^*), \cr
\gamma_i \to 
\exp\left [N \left ( (1-\alpha)\log 2 + \alpha H (x) 
\right )\right ]& (x < x^*), 
\end{array}
\right .
\label{RS2_ni_average} 
\end{eqnarray}
where $x^*$ is determined by Eq. (\ref{x*}) and
\begin{eqnarray}
\omega^{i \sum_{\mu =1}^n \delta_{i,i_\mu } }
&= & \exp \left [ -n \beta i  \right ]
\to \exp \left [ -n N \alpha \beta x \right ], 
\label{RS_2occupied_contribution}
\end{eqnarray}
hold for large $N$. 
Notice that $\left \langle n_i^{\sum_{\mu=1}^n 
\delta_{i,i_\mu}} \right \rangle$ behaves differently
depending on whether $x>x^*$ or not. 
Therefore, the behavior of the dominant contribution 
obtained by maximizing 
$\left \langle n_i^{\sum_{\mu=1}^n \delta_{i,i_\mu}} 
\right \rangle \omega^{i \sum_{\mu =1}^n \delta_{i,i_\mu } }$
is different depending on the relation between $x^*$
and the position of the maximum $x_c$. 
%%%%%%%%%%%%%%%%
For $\beta < \beta_c$ 
or $\alpha \le 1$, $x_c=(1-\tanh\frac{\beta}{2})/2$ becomes 
greater than $x^*$, which yields the RS1 solution, Eq. (\ref{rs1}), 
in spite that the RS2 ansatz is currently employed. 
This is because the RS2 ansatz for {\em energy levels }
does not necessarily mean that $n$ replica {\em states}
are identical, in particular, when the occupation number 
$n_i$ is exponentially large, to which $x_c > x^*$ corresponds. 
In such cases, even if $n$ replica states are allocated to 
an identical energy level, they are typically distributed to 
$n$ different states placed at the energy level, 
which corresponds to the RS1 ansatz of CDREM, and, therefore, 
Eq. (\ref{rs1}) should be reproduced. 
On the other hand, for $\beta > \beta_c$, 
$\left \langle n_i^{\sum_{\mu=1}^n \delta_{i,i_\mu}} 
\right \rangle \omega^{i \sum_{\mu =1}^n \delta_{i,i_\mu } }$
is maximized at $x_c=(1-\tanh\frac{n\beta}{2})/2 < x^*$. 
This provides 
\begin{eqnarray}
\left \langle Z^n \right \rangle
= \exp N \left (\log 2+ \alpha \log \left ( \cosh \frac{n \beta }{2} \right )
\right ),
\label{RS2_appendix}
\end{eqnarray}
which is identical to the RS2 solution of CDREM, Eq. (\ref{rs2}), 
and equivalent to Eq. (\ref{rs2_ours}). 
As Eq. (\ref{rs2}) can be dominant at $n=1,2,\ldots$ 
only for $\beta > \beta_c$, this is consistent with 
the result of the replica analysis of CDREM. 

\item{1RSB}\\
In 1RSB, $n$ replica levels are assumed to be equally allocated
to $n/m$ levels. 
Therefore, for a given level $i$, 
\begin{eqnarray}
\sum_{\mu=1}^n \delta_{i,i_\mu}=
\left \{
\begin{array}{ll}
m & (\mbox{when $i$ is the $n/m$ allocated energy levels}), \cr
0 & (\mbox{otherwise}).
\end{array}
\right .
\end{eqnarray}
When $i$ is the allocated level, 
\begin{eqnarray}
\left \langle n_i^{\sum_{\mu=1}^n \delta_{i,i_\mu}} 
\right \rangle 
&\sim &
\left \{
\begin{array}{ll}
\gamma_i^m 
\to \exp\left [m N \left ( (1-\alpha)\log 2 + \alpha H (x) 
\right )\right ]& (x > x^*), \cr
\gamma_i \to 
\exp\left [N \left ( (1-\alpha)\log 2 + \alpha H (x) 
\right )\right ]& (x < x^*), 
\end{array}
\right .
\end{eqnarray}
and
\begin{eqnarray}
\omega^{i \sum_{\mu =1}^n \delta_{i,i_\mu } }
&= & \exp \left [ -m \beta i  \right ]
\to \exp \left [ -m N \alpha \beta x \right ], 
\end{eqnarray}
hold for large $N$. 
Similarly for the case of RS2, this reproduces the RS1 solution, 
Eq. (\ref{rs1}), for $\beta < \beta_c$ or $\alpha \le 1$.  
So, we focus on the low temperature phase $\beta > \beta_c$. 
Then, $\left \langle n_i^{\sum_{\mu=1}^n \delta_{i,i_\mu}} 
\right \rangle \omega^{i \sum_{\mu =1}^n \delta_{i,i_\mu } }$
is maximized at $x_c=(1-\tanh\frac{m \beta}{2})/2 < x^*$ to 
\begin{eqnarray}
\exp  N\left ( \log 2+\alpha \log \left (
\cosh\frac{m \beta}{2} \right ) - \frac{m \alpha\beta}{2} \right ) ,
\label{first_level}
\end{eqnarray} 
which represents the contribution from one of the $n/m$ allocated levels. 
The number of ways to select $n/m$ out of $M$ levels equally 
allocating $n$ replicas is negligible for subsequent calculation. 
Taking contributions from all the $n/m$ allocated levels
and extremization with respect to $m$ into account
offers 
\begin{eqnarray}
\left \langle 
Z^n \right \rangle
&\sim &
\exp N \left (
\mathop{\rm extr}_{m}
\left \{
\frac{n}{m} \left (\log 2 
+\alpha \log \left (
\cosh\frac{m \beta}{2} \right ) \right )
\right \}
\right ) \cr
&=&
\exp  N n \alpha \beta \left (\frac{1}{2}-x^*\right )\cr
&=& \exp N\left (\frac{n\alpha \beta}{2}\tanh \frac{\beta_c}{2} \right ), 
\label{1RSB}
\end{eqnarray}
which is identical to the 1RSB solution of CDREM, 
Eq. (\ref{1rsb}), and equivalent to Eq. (\ref{1rsb_ours}). 
As Eq. (\ref{1rsb}) can be dominant $n=1,2,\ldots$ 
only for $\beta > \beta_c$, this is consistent 
with the result of the replica analysis of CDREM. 
\end{itemize}

\section{Efficient sampling in CDREM}
\label{CDREM} %Empty argument \section{} yields `Appendix'. 
%Fast sampling of a set of $n_i$ like that in 
We here show that 
one can sample $\{n_i \}$ with an $O(M)$ computational cost
in CDREM as well as in GCDREM.  According to the probability 
distribution for each level
\begin{eqnarray}
P(E_i)=2^{-M}
\left(
\begin{array}{c}
M\\
\frac{1}{2}M+E_i
\end{array}
\right),\ \ \left (E_i=i-\frac{M}{2} \right ),
\end{eqnarray}
the probability to sample a configuration $(n_0,n_1,\cdots,n_M)$ is
\begin{eqnarray}
{\cal P}(n_0,n_1,\cdots,n_M)
=\{P(E_0)\}^{n_0}\{P(E_1)\}^{n_1}\cdots \{P(E_M)\}^{n_M}\frac{2^N!}{n_0!n_1!\cdots n_M!}.
\end{eqnarray}
To generate a sample $(n_0,n_1,\cdots,n_M)$ in practice, 
we first determine $n_0$ according to the probability
\begin{eqnarray}
{\cal P}(n_0,arbitrary)
=(p_0)^{n_0}(1-p_0)^{2^N-n_0}\frac{2^N!}{n_0!(2^N-n_0)!}\\
(p_0\equiv P(E_0)).
\end{eqnarray}
We then determine $n_1$ according to the probability
\begin{eqnarray}
{\cal P}(n_0;n_1,arbitrary)
=(p_1)^{n_1}(1-p_1)^{2^N-n_0-n_1}\frac{2^N!}{n_1!(2^N-n_0-n_1)!}\\
\left(p_1\equiv \frac{P(E_1)}{1-P(E_0)}\right).
\end{eqnarray}
Repeating this procedure to $n_{M-1}$ as
\begin{eqnarray}
{\cal P}(n_0,n_1,\cdots;n_{M-1},n_M)
=(p_{M-1})^{n_1}(1-p_{M-1})^{2^N-\sum_{i=0}^{M-1} n_i }\frac{2^N!}{n_{M-1}!(2^N-\sum_{i=0}^{M-1} n_i )!}\nonumber\\
\left(p_{M-1}\equiv \frac{P(E_{M-1})}{1-\sum_{i=0}^{M-2}P(E_{i})}\right),\nonumber\\
\end{eqnarray}
we obtain a set of $(n_0,n_1,\cdots,n_M)$. 
This guarantees that the identity
\begin{eqnarray}
\sum_{i=0}^{M} n_i = 2^N, 
\end{eqnarray}
holds, which characterizes CDREM. Similarly for the case of GCDREM, 
this can be performed in $O(M)$ computation. 

\begin{figure}
       \centerline{\includegraphics[width=12cm]
                                   {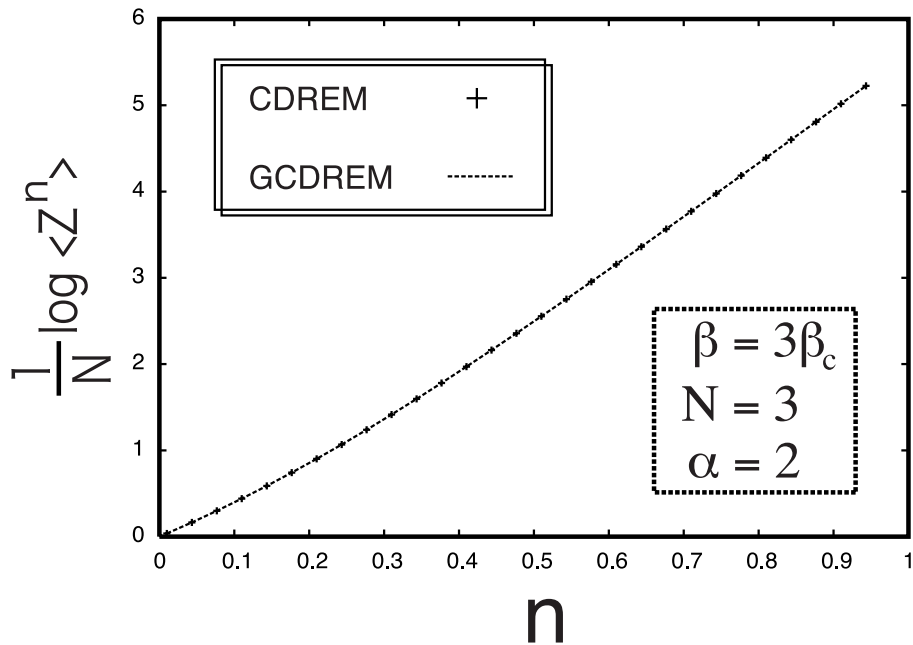}}
\caption{The free energies obtained from the sets $\{n_i\}$ using CDREM
and GCDREM. We find that CDREM and GCDREM give
almost same results even for $N=3$.}
\label{REM}
\end{figure}

A comparison of numerically evaluated moments of the partition 
function between CDREM and GCDREM is presented in Fig.~\ref{REM}.  
This indicates that the difference between the two models 
is almost indistinguishable even for $N=3$. 
Since the consistency becomes better as $N$ grows larger\cite{Mou1,Mou2}, 
working in GCDREM instead of CDREM is justified when $N$ is large. 

%
%\section{Second Appendix}

\end{document}